\documentclass{appolb}
\usepackage{graphicx}
\def\Eq#1{Eq.~(\ref{#1})}
\begin{document}
% \eqsec  % uncomment this line to get equations numbered by (sec.num)
\title{Shear viscosity in QCD and why it's hard to calculate
\thanks{Presented at Criticality 2020}%
% you can use '\\' to break lines
}
\author{Guy D.~Moore
\address{Institut f\"ur Kernphysik, Technische Universit\"at
  Darmstadt, Schlossgartenstra{\ss}e 2, 64289 Darmstadt, Germany}
}
\maketitle
\begin{abstract}
Shear viscosity is a dynamical property of fluid systems close to
equilibrium, describing resistance to sheared flow.  After reviewing
the physics of viscosity and the reason it is usually difficult to
compute, I discuss its importance within the theory of QCD and the
obstacles to carrying out such a computation.  A diagrammatic analysis
requires extensive resummations and even then convergence is poor at
physically relevant couplings.  Lattice approaches require a poorly
controlled analytical continuation of data from the Euclidean to the
Minkowski domain.  At present our best results for QCD shear viscosity
come from the hydrodynamical interpretations of experiments, with
first-principles calculations trailing behind.
\end{abstract}
%\PACS{PACS numbers come here}
  
\section{Introduction}

The topic of this conference is the thermodynamics of the Quark-Gluon
Plasma.  This talk won't quite be about that -- it will be
``thermodynamics adjacent,'' talking about something closely related.
Specifically, viscosity is a property of fluids (anything without
crystallization, that is, long-range ordering and spontaneous rotation
symmetry breaking, so the medium is free to move around with no
long-term ``memory'' of its previous form), when the fluid is close to
equilibrium.  Technically we can relate it, by
fluctuation-dissipation, to a fluid \textsl{in} equilibrium, but its
unequal-time correlation functions which are technically beyond what
people usually mean when they say ``thermodynamics.''  So it's
``thermodynamics adjacent'' in the sense that you can talk about it
for an equilibrium system, but it's not strictly a thermodynamical
property.

Viscosity plays a big role in the development of real systems.  In the
case of the Quark-Gluon Plasma we are talking about heavy ion
collisions, where viscosity is important in the hydrodynamic
development of the system \cite{Schafer:2009dj,Teaney:2009qa}.  It
might also be important in
the biggest heavy-ion collisions of all, the neutron star mergers
\cite{NSM}.  My goal in this talk is to review what shear viscosity
is, and to lay out what we do and don't know about it in the case of
Quantum Chromodynamics and the Quark-Gluon Plasma.  This is only a
review of the established literature; nothing in this talk will be
new, but I hope it will give a nice perspective and introduction to
the topic, which is what the conference organizers requested from me.

In the next section, I will remind the reader of the physical meaning,
and then of the definition, of shear viscosity.  Then I will explain
why, for many systems, it's really not easy to compute it from first
principles.  In Section \ref{sec:pertthy} I will show that for QCD at
weak coupling, we can compute it analytically; but the computation
requires very extensive resummations of the naive perturbative
expansion, and the convergence of the series leaves a lot to be
desired.  Then Section \ref{sec:latt} reviews attempts to get at the
shear viscosity nonperturbatively from the lattice.  Finally, I give a
brief summary in the conclusions.

\section{What is viscosity?}
\label{sec:whatis}

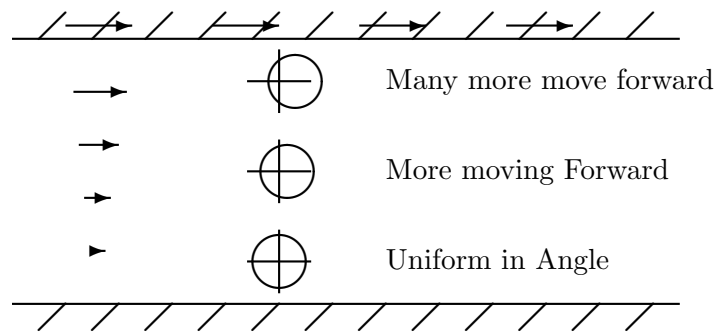
\begin{figure}[ht]
  \centerline{
    \begin{picture}(260,140)
	\thicklines
	\put(00,10){\line(1,0){250}}
	\put(00,110){\line(1,0){250}}
	\multiput(30,30)(55,0){1}{\vector(1,0){5}}
	\multiput(27,50)(55,0){1}{\vector(1,0){10}}
	\multiput(25,70)(55,0){1}{\vector(1,0){15}}
	\multiput(23,90)(55,0){1}{\vector(1,0){20}}
	\multiput(20,115)(55,0){4}{\vector(1,0){25}}
	\multiput(10,110)(20,0){12}{\line(1,1){10}}
	\multiput(10,0)(20,0){12}{\line(1,1){10}}
	\put(100,14){\line(0,1){24}}
	\put(88,26){\line(1,0){24}}
	\put(100,26){\circle{20}}
	\put(140,23){Uniform in Angle}
	\put(100,48){\line(0,1){24}}
	\put(88,60){\line(1,0){24}}
	\put(103,60){\circle{20}}
	\put(140,57){More moving Forward}
	\put(100,82){\line(0,1){24}}
	\put(88,94){\line(1,0){24}}
	\put(106,94){\circle{20}}
	\put(140,91){Many more move forward}
  \end{picture}}
\caption{Gas between two plates, with the top plate moving forward and
  the bottom plate at rest.  The velocity distribution is hinted at
  the left; the microscopic distribution of particle-number with
  direction is illustrated to the right.}\label{fig1}
\end{figure}

To illustrate what shear viscosity is about, think about a gas trapped
between two plates.  The plate at the top is moving with a constant
velocity $v_0$; a distance $L$ below it, the lower plate is at rest.
The fluid then develops a space-dependent velocity
$v(z) = v_0 z/L$, and the individual atoms develop a space-dependent
directional distribution, as illustrated in Figure \ref{fig1}.

\begin{figure}
  \centerline{
    \begin{picture}(300,205)
	\thicklines
	\put(10,20){\line(1,0){280}}
	\put(10,200){\line(1,0){280}}
	\multiput(20,10)(25,0){11}{\line(1,1){10}}
	\multiput(20,200)(25,0){11}{\line(1,1){10}}
	\put(0,107){Initial}
	\put(40,50){\line(1,0){50}}
	\put(65,25){\line(0,1){50}}
	\put(65,50){\oval(42,42)[t]}
	\put(65,50){\oval(38,38)[t]}
	\put(65,50){\oval(40,40)[b]}
        \put(150,180){{\small{note: longer free path}}}
        \put(150,167){{\small{means more anisotropic, so}}}
        \put(150,154){{\small{larger viscosity.}}}
	\put(40,110){\line(1,0){50}}
	\put(65,85){\line(0,1){50}}
	\put(69,110){\circle{40}}
	\put(40,170){\line(1,0){50}}
	\put(65,145){\line(0,1){50}}
	\put(73,170){\oval(40,40)[t]}
	\put(73,170){\oval(38,38)[b]}
	\put(73,170){\oval(42,42)[b]}
	\put(85,70){\line(1,1){50}}
	\put(135,120){\vector(1,0){40}}
	\put(85,150){\line(1,-1){50}}
	\put(135,100){\vector(1,0){40}}
	\put(180,110){\line(1,0){50}}
	\put(205,85){\line(0,1){50}}
	\put(205,110){\oval(42,42)[t]}
	\put(205,110){\oval(38,38)[t]}
	\put(213,110){\oval(42,42)[b]}
	\put(213,110){\oval(38,38)[b]}
	\put(240,107){Final}
	\put(40,0){Skewed momentum distribution.}
  \end{picture}}
  \caption{Velocity distribution in more detail, remembering that
    molecules have a finite mean free path.  Free propagation and
    space-dependent velocity lead to a locally anisotropic velocity
    distribution.}\label{fig2}
\end{figure}
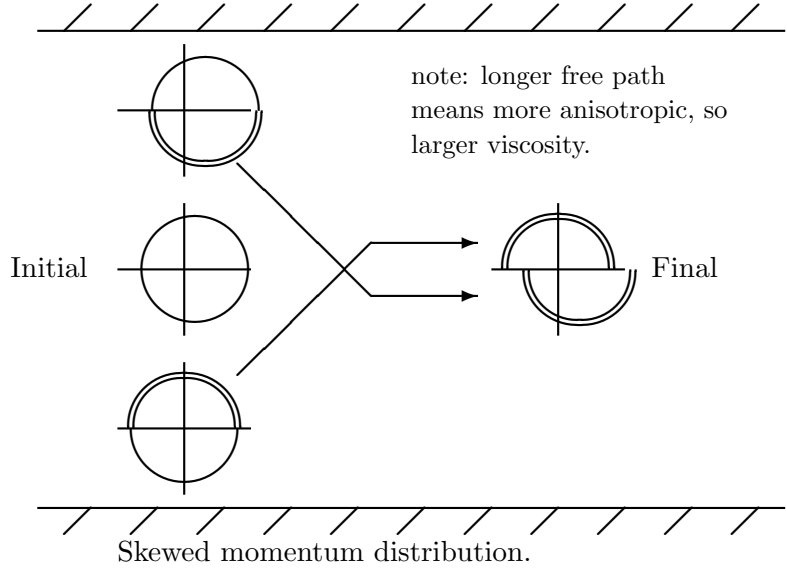

Now remember that this is a gas; the individual molecules are free to
fly, and will typically fly a distance $\lambda$ (the mean free path)
between scatterings with other gas molecules.  That means that the
molecules at any location actually came from somewhere else.  At the
center of the gas, the downward-moving molecules came from higher up,
and the upward-moving molecules came from lower down.  Since the
molecules above this point tend to move forward, the downward-moving
molecules tend to move forward; but the upward-moving molecules come
from somewhere with a smaller forward velocity, and they are less
likely to move forward.  Therefore, there is a correlation between
transverse and vertical velocity in the middle, as illustrated in
Figure~\ref{fig2}.  The same correlation occurs everywhere in the
fluid.

The particles which strike the bottom plate tend to move forward, and
therefore impart a forward-pointing force on that plate.  At the top,
the upward-going particles are moving forward at a smaller velocity
than the plate, and impart a net backwards force.  This force is
proportional to the plate area $A$ and to $v_0/L$ the velocity
gradient; we can write
%\begin{equation}
%  \label{etadef}
$F = \eta \frac{A v_0}{L}$ or $T_{zx} = -\eta \partial_z v_x$,
%\end{equation}
where $\eta$ is some material-dependent coefficient, called the
\textsl{shear viscosity}.
In the last step we recalled that a force per area is the
definition of the stress tensor, and $z$ and $x$ are the directions of
the surface-normal and the force respectively.  More generally, for a
fluid undergoing generalized inhomogeneous flow, we can write
\begin{equation}
\label{shearbulk}
  T_{ij} - T_{ij,\mathrm{eq}} =
  - \eta \left( \partial_i v_j + \partial_j v_i
  - \frac 23 \delta_{ij} \partial_k v_k \right)
  - \zeta \delta_{ij} \partial_k v_k \,.
\end{equation}
The $\delta_{ij} \partial_k v_k$ term separates the pure-divergence
part of the flow, which has an independent coefficient $\zeta$ called
the bulk viscosity, which I won't discuss.

%A few points are worth emphasizing.
%First,
$\eta$ is by definition a
linearization about equilibrium; we assume $\partial_i v_j$ is
``small'' in some sense -- generally it is enough that the resulting
$T_{ij}$ is small compared to the pressure.  Also, we are assuming
that the gradients don't change rapidly, that is, we assume a
hierarchy in which
$\lambda \partial_i v_j \ll \lambda^2 \partial_i \partial_j v_k$.
If one or the other condition does not apply, shear viscosity -- and
hydrodynamics -- is generally not a useful concept.

One last point:  the fluid flow $\partial_i v_j$ is itself related to
the stress tensor $T_{ij}$.  For instance, if we had a system in
perfect equilibrium, we could generate fluid flow with nonvanishing
$\partial_i v_j$ by \textsl{distorting the metric}
$g_{\mu\nu} = \eta_{\mu\nu} + h_{\mu\nu}$, and in the path integral
this would arise from an interaction Hamiltonian of
$H_I = h_{\mu\nu} T^{\mu\nu}$.  Following this logic (see
\cite{Kubopaper}), one can derive a Kubo relation for the shear
viscosity in terms of equilibrium, unequal time, retarded correlation
functions:
\begin{equation}
  \label{kubo}
  \eta = i \partial_\omega \int d^3 x \int_0^\infty dt
  \: e^{i\omega t}
  \left. \left\langle \Big[ T^{xy}(x,t)\,,\, T^{xy}(0,0) \Big]
  \right\rangle \right|_{\omega=0} \,.
\end{equation}
We will need this relation repeatedly in what follows.

\section{Why is it hard to compute?}
\label{sec:challenge}

What is the shear viscosity of air at STP?  What about water?
These sound like innocent questions, but answering them is hard.
Naively, we should be able to:  we know the theory describing each
system.  It's QED with electrons, oxygen nuclei, nitrogen nuclei, and
hydrogen nuclei as the particle species.  The theory has a weak
coupling $\alpha_{\mathrm{EM}} = 1/137$ and it's nonrelativistic, so
it \textsl{sounds naively} like this is a regime where we can do the
calculation.  But that manifestly doesn't work.  The problem is hard
because
\begin{itemize}
\item
  The theory is not actually weakly coupled in a useful way; potential
  energies are of order kinetic energies which are of order the
  temperature, while perturbation theory generally works by perturbing
  in potential energies being smaller than the other two.
\item
  It's a many-body problem.  More is different \cite{moreisdifferent}.
\item
  We are after a low-frequency and long-distance property.
\end{itemize}
To see the challenge, think first of air.  This is a system of weakly
coupled, well separated diatomic molecules.  If we could compute the
differential scattering cross-sections between molecules, we could
write a Boltzmann equation which could solve for the viscosity.  But
in practice molecular scattering is a fairly hard atomic and molecular
physics problem; first-principles calculations are challenging, and
they are made more so because we have to account for the rotational
states of the molecules.  But if we had scattering data, we could
predict the viscosity without too much effort.

The case of water is harder, because the molecules are all pressed
against each other in a highly correlated system.  No amount of
molecule-on-molecule scattering data will tell us how to deal with
this system, and in the end the most efficient thing to do is simply
to measure the viscosity and give up on a first-principles
calculation.

We might worry that the same kind of complexity hits us in QCD.  At
physically interesting temperatures $T < 1\,\mathrm{GeV}$, the
coupling is not small, $\alpha_s \simeq 0.3$, so there will be strong
correlations and we don't have a good perturbative expansion.  And the
theory is fully relativistic, so it's not clear that particles or
particle number are useful concepts.  So to begin with let's ask a
simpler question:  what happens at high temperatures where the
coupling is actually weak?

\section{Weak coupling:  resummations and series convergence}
\label{sec:pertthy}

The case of high-temperature QCD, where the coupling is small, should
be more tractable.  Weak coupling means that there are long-lived
quasiparticles, and the behavior is loosely analogous to a gas like
the air.  But unlike molecular gases, we have an efficient
tool for computing scattering cross-sections, perturbation theory.  So
this case should be tractable.

\begin{figure}
  \includegraphics[width=0.44\textwidth]{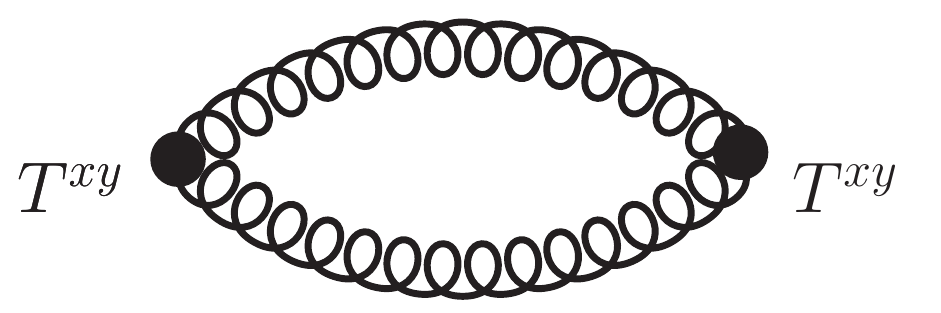}
  \hfill
  \includegraphics[width=0.5\textwidth]{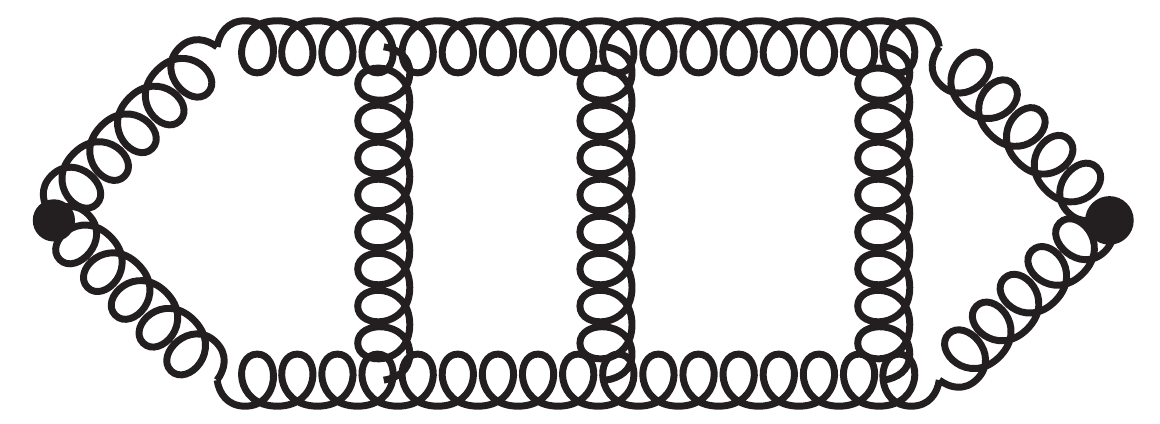}
  \caption{Left:  the naive leading-order diagram.
    Right:  an example of the family of diagrams which actually
    determines the \textsl{leading-log} viscosity.}
  \label{fig3}
\end{figure}

We want to evaluate the correlation function shown in \Eq{kubo}.
Naively, at leading order we should only need to calculate the
leftmost diagram in Figure~\ref{fig3}.  And if we were computing at
generic off-shell $(\omega,p)$ with $|\omega^2 - p^2| \sim T^2$, this
would be correct.  But we don't want that case; we want $p=0$ and the
small-$\omega$ limiting behavior, specifically the linear-in-$\omega$
term at $\omega=0$.  If the external momentum inserted by the $T^{xy}$
operator is almost zero, then the upper propagator is on shell exactly
where the lower propagator is; pinching poles.  Computing this
correctly requires including self-energies, to push the poles off the
real frequency axis.  The result is that there is an
$\mathcal{O}(1/\alpha_s)$ enhancement.  But inserting a vertical line,
as in the right diagram in Figure \ref{fig3}, gives another pair of
lines which are on shell, and another $1/\alpha_s$ enhancement.  Each
vertical line costs $\alpha_s$ for the vertices, but gives another
$1/\alpha_s$ enhancement, so we need to resum any number of such
diagrams; for a detailed discussion see
\cite{Jeon,AMY1,AMY6}.  The short version is that such diagrams,
together with a ``pinching pole'' approximation in which we expand
about the two propagators being nearly on-shell, resum into kinetic
theory.

\begin{figure}[th]
  \centerline{
    \includegraphics[width=0.7\textwidth]{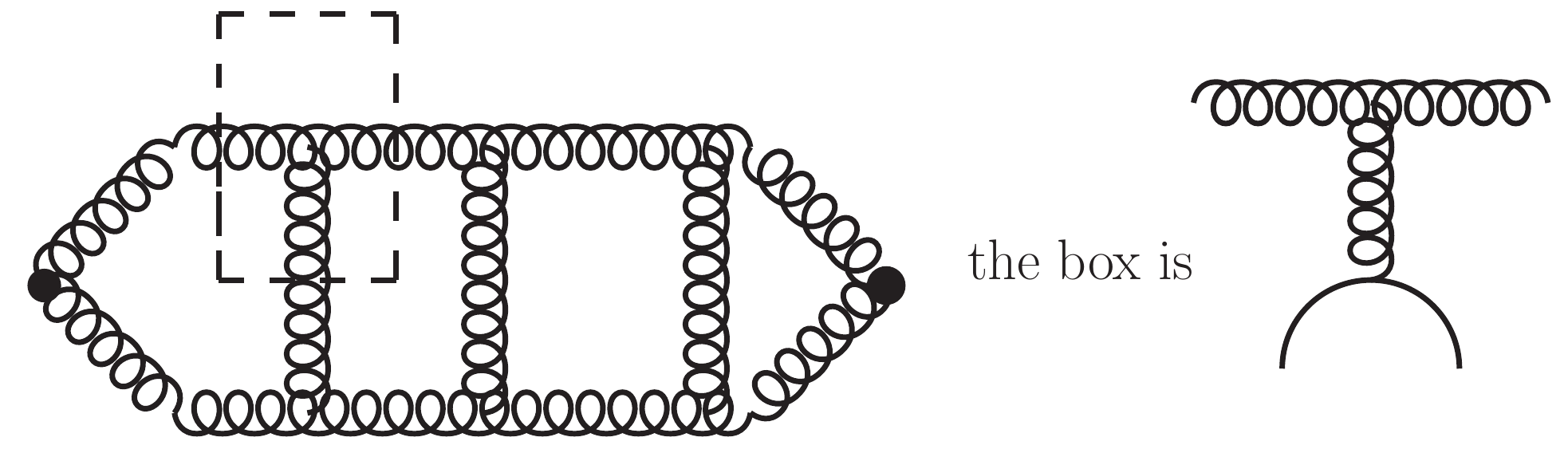}}
  \caption{Illustration of what kinetic theory has to do with ladder
    graphs.  The indicated part of the ladder graph corresponds to the
    scattering matrix element shown; the lower half of the rung is the
    starred matrix element.  The propagator must always be ``cut''
    through a self-energy insertion, corresponding to a second
    particle against which the particle in consideration scatters.}
  \label{fig4}
\end{figure}

\begin{figure}[th]
  \centerline{
    \includegraphics[width=0.6\textwidth]{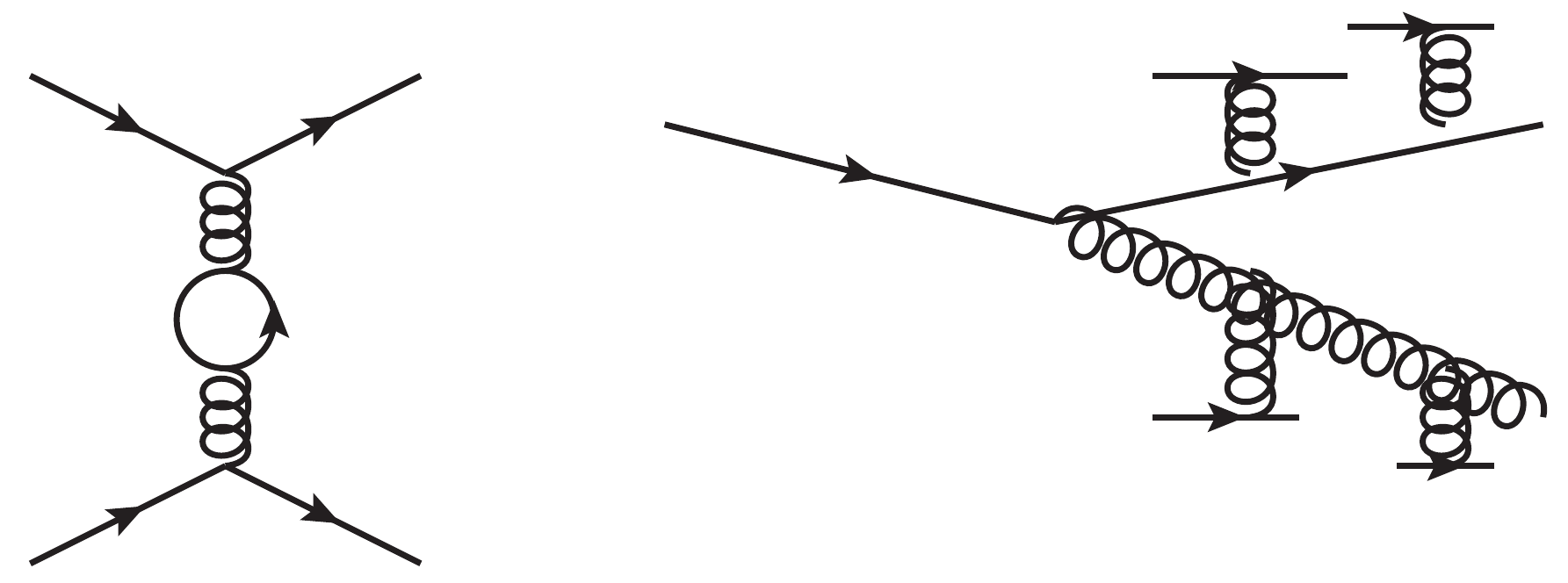}}
  \caption{Other relevant scattering processes.  Left:  normal
    scattering requires ``HTL'' resummations on propagators.  Right:
    inelastic splitting with multiple scatterings must also be
    accounted for.}
  \label{fig5}
\end{figure}

But the story is still more complex.  Even at leading-logarithmic
order \cite{AMY1} it is necessary to include self-energy corrections
on low-momentum gluonic lines, the so-called Hard Thermal Loop
resummation \cite{HTL}, see Figure~\ref{fig4}.  And at leading order
\cite{AMY6} one needs to include a \textsl{much} more complex set of
processes; inelastic splitting, induced by any number of scatterings,
correctly accounting for the Landau-Pomeranchuk-Migdal destructive
interference effect \cite{LP,M}, see Figure~\ref{fig5}.  For a (much)
more detailed exposition on this issue, see the original literature
\cite{AMY2,AMY5,AMY6}.

\begin{figure}[th]
  \centerline{
    \includegraphics[width=0.6\textwidth]{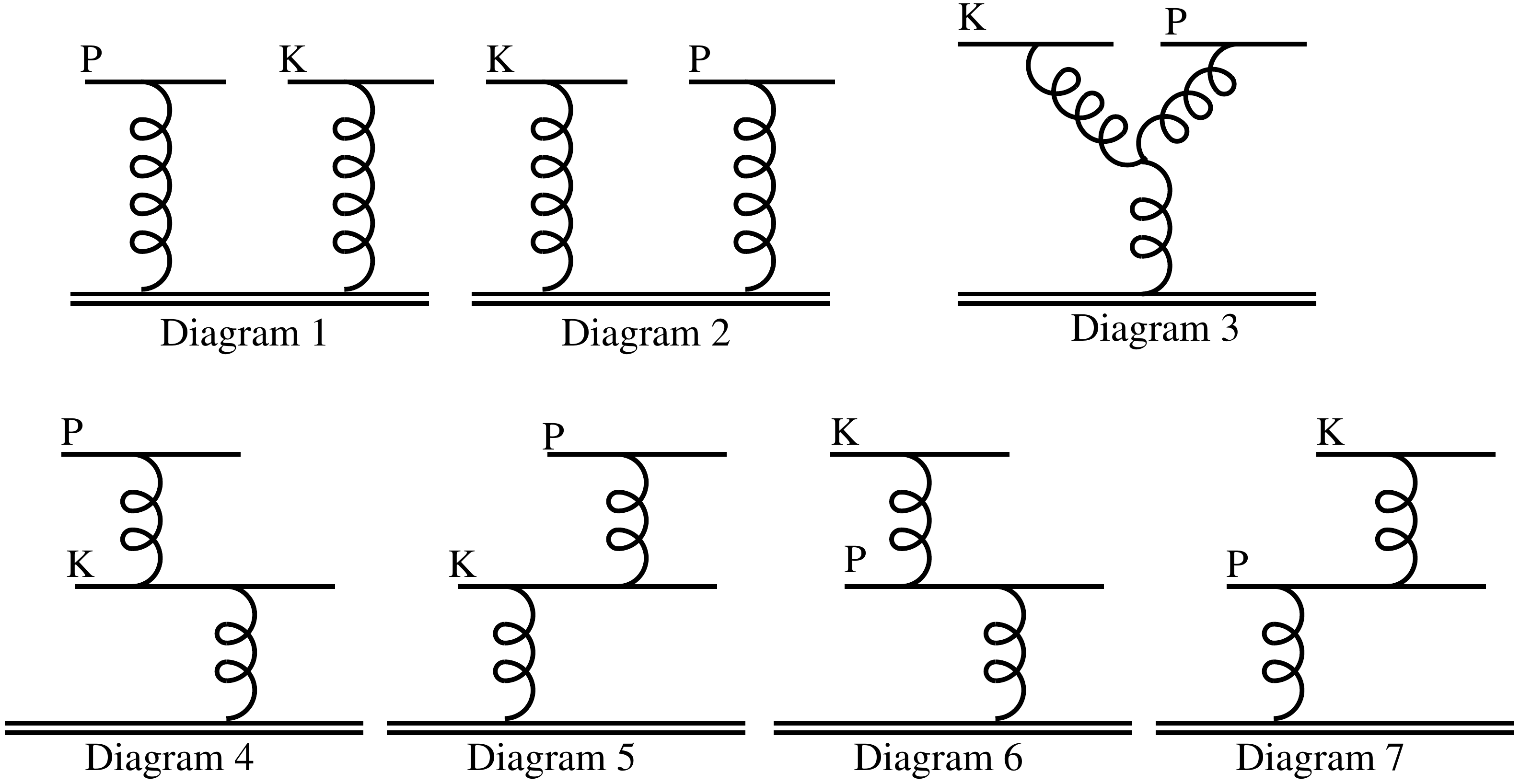}}
  \caption{The types of effect which arise at next-to-leading order.
    Two successive scatterings overlap in time, so the interference
    between various orderings, connections, and nonabelian effects
    must be included.} \label{fig6}
\end{figure}

\begin{figure}[th]
\begin{minipage}{0.5\textwidth}
  \includegraphics[width=0.98\textwidth]{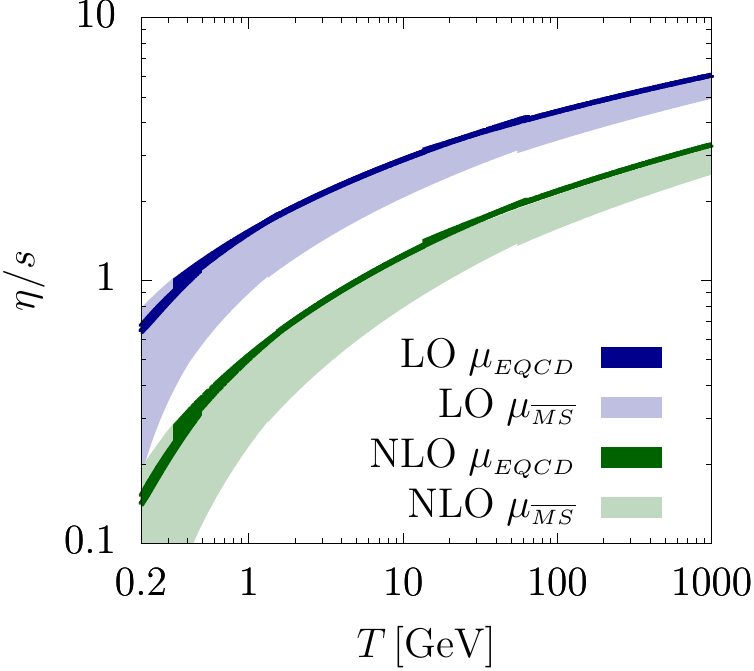}
\end{minipage} \hfill
\begin{minipage}{0.45\textwidth}
  \caption{Shear viscosity to entropy density ratio at LO and NLO in
    QCD as a function of temperature.  The difference between orders
    is already a factor of 2 at $T = 100\:\mathrm{GeV}$.}
  \end{minipage}
  \label{fig7}
\end{figure}

What about next-to-leading order?  At this order, successive
scatterings temporally overlap and interfere with each other, creating
complex effects.  But because the involved particles move at close to
the speed of light, there are resummation methods which can be
applied, rendering these NLO effects calculable
\cite{Simon,jetpaper,NLOpaper}.
The first interesting result is that, contrary to vacuum perturbation
theory, the first corrections do not appear at
$\mathcal{O}(\alpha_s)$; they occur at
$\mathcal{O}(\sqrt{\alpha_s})$.  Second, the effects are surprisingly
large.  A comparison of the leading-order and next-to-leading order
(LO and NLO) results for the shear viscosity are shown in
Figure~\ref{fig7}.

The result is that perturbation theory for the shear viscosity only
converges at temperatures well above 100 GeV.

\section{What about determining it from the lattice?}
\label{sec:latt}

So perturbation theory does not work well.  What about a
first-principles, nonperturbative approach, Lattice QCD?  The
challenge is that the lattice deals with correlation functions in
Euclidean time, and \Eq{kubo} is written in terms of Minkowski time.
Also, \Eq{kubo} refers to a commutator of operators.  The lattice
treats the path integral
\begin{equation}
  \label{Eucl}
Z = \int \mathcal{D}(A^\mu,\bar\psi,\psi) e^{-S_E[A,\bar\psi,\psi]}
\end{equation}
which converges absolutely and has a positive integrand.  That is what
makes numerical integration possible.  But the Minkowski correlation
functions are determined by the path integral
\begin{equation}
  \label{Mink}
  Z = \int \mathcal{D}(A^\mu,\bar\psi,\psi) e^{iS_M[A,\bar\psi,\psi]}
\end{equation}
which does not converge absolutely and has extremely important phases
and phase cancellations.  This (so far) makes Minkowski lattice
techniques unavailable.  But computing a correlation function from the
first path integral gives us
$G(\tau) \equiv \langle T^{xy}(x,i\tau) T^{xy}(0,0) \rangle$, that is, the
correlation function at imaginary time.  This isn't the correlator we
want, but at least it is analytically related to our correlator.
Generalizing \Eq{kubo} slightly, we can define
\begin{equation}
\sigma(\omega) =  i \int d^3 x \int_{-\infty}^\infty dt \; e^{i\omega t}
\langle [ T^{xy}(x,t) , T^{xy}(0,0) ] \rangle \,, \qquad
\eta = \lim_{\omega \to 0} \frac{\sigma(\omega)}{\omega} \,.
\end{equation}
And this is related to the correlator we can measure via
\begin{equation}
  G(\tau) = \int \frac{d\omega}{2\pi} \frac{\sigma(\omega)}{\omega}
  K(\omega,\tau) \,, \qquad
  K(\omega,\tau) = \frac{\omega \mathrm{cosh}(\omega(\tau{-}1/2T))}
  {\sinh(\omega/2T)}
\end{equation}
This represents an ``inverse problem.''  We measure $G(\tau)$ at a
series of $\tau$ values, and it tells us a series of integrals of
$\sigma(\omega)$, each with different integral kernels, and with error
bars.  From this limited and incomplete information, we need to
reconstruct the original $\sigma(\omega)$.

\begin{figure}[th]
\begin{minipage}{0.53\textwidth}
    \includegraphics[width=0.99\textwidth]{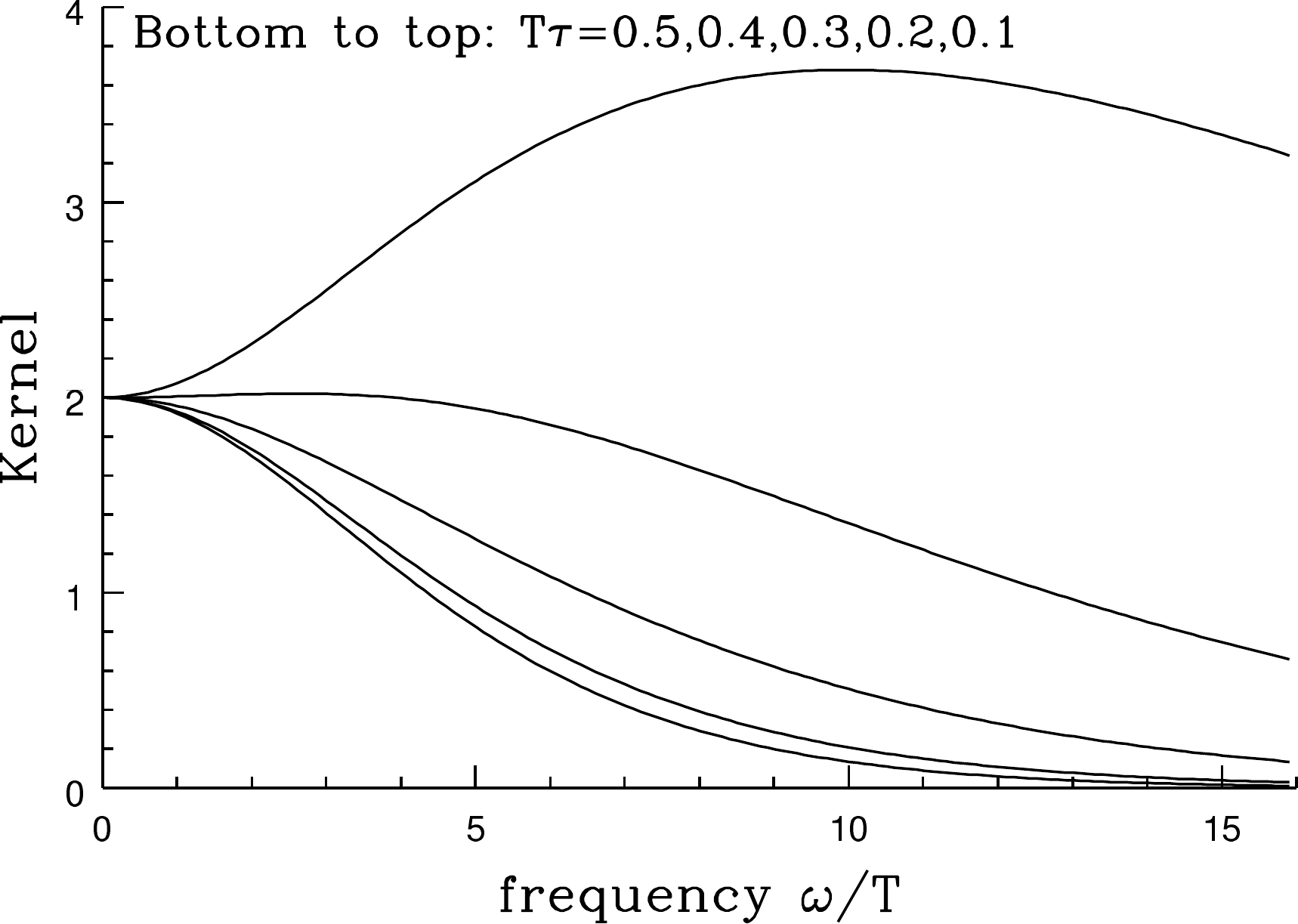}
\end{minipage} \hfill
\begin{minipage}{0.44\textwidth}
    \caption{The integral kernel
      $\omega \cosh(\omega(\tau-1/2T))/\sinh(\omega/2T)$
      for several $\tau$ values.}
\end{minipage}
\label{fig8}
\end{figure}

\begin{figure}
  \includegraphics[width=0.46\textwidth]{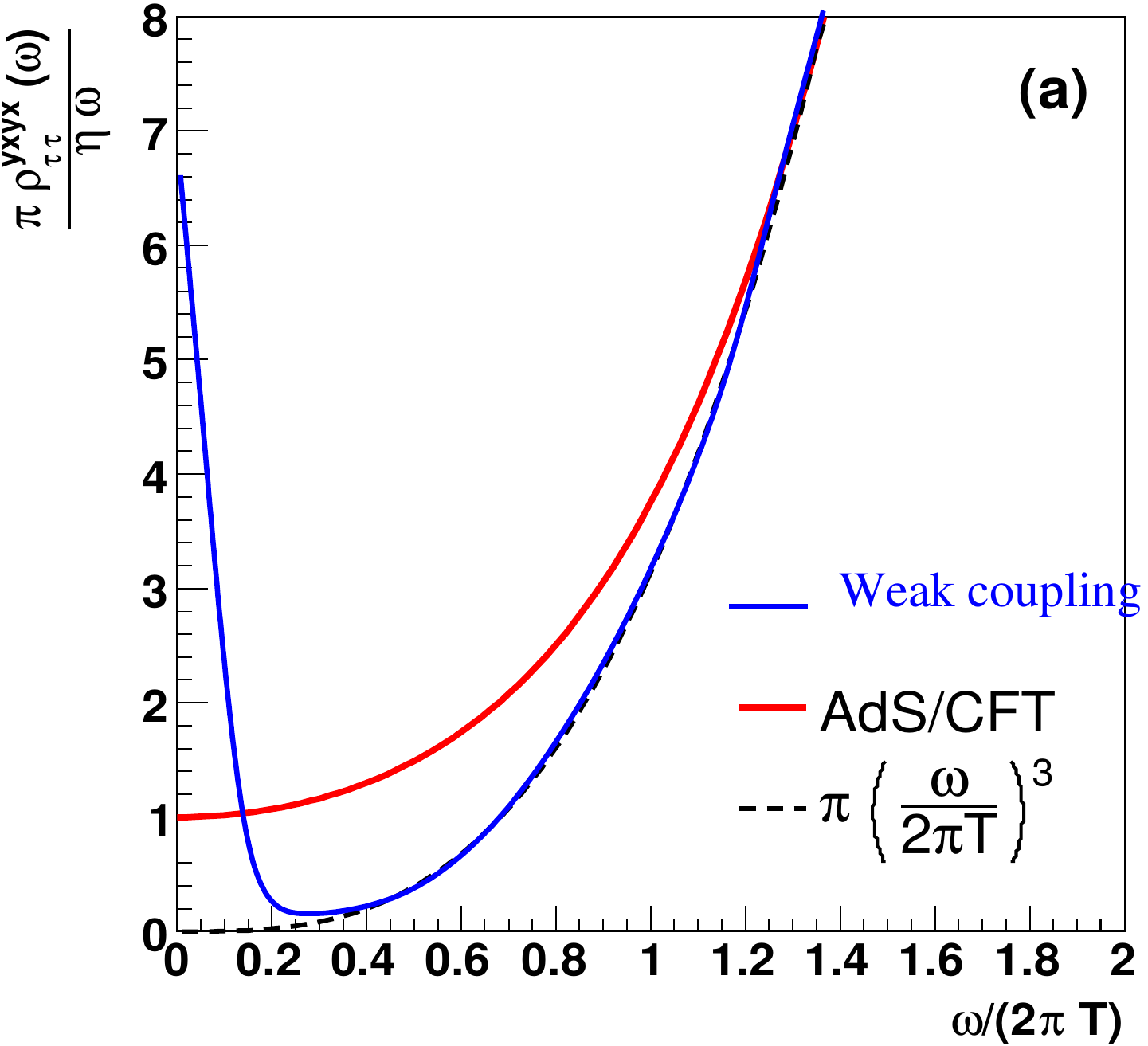}
  \hfill
  \includegraphics[width=0.46\textwidth]{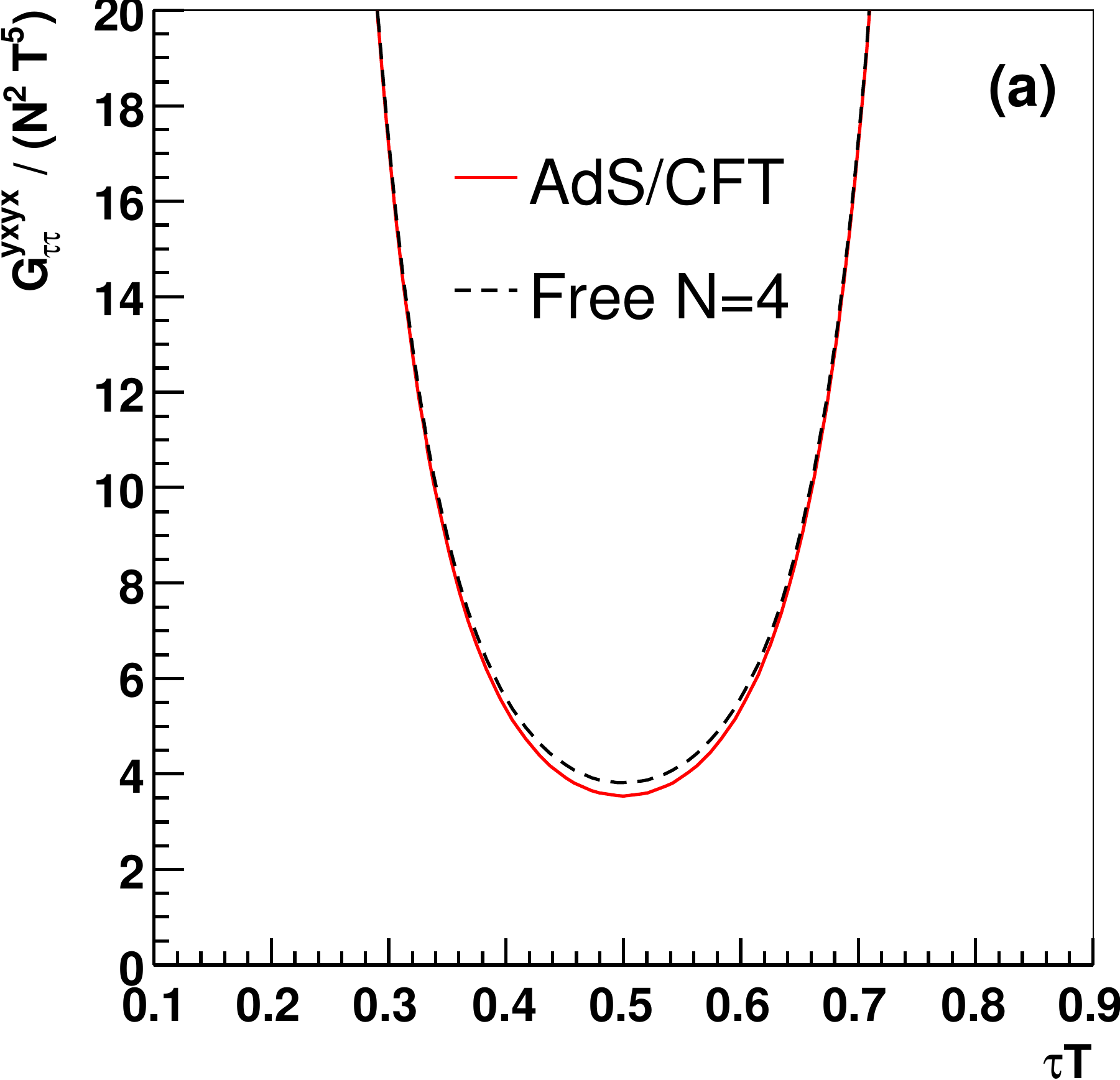}
  \caption{Left:  the (very different) spectral functions
    $\sigma(\omega)/\omega)$ for two theories, weakly-coupled QCD and
    strongly-coupled $\mathcal{N}=4$ SUSY QCD.  Right:  their almost
    identical Euclidean correlation functions $G(\tau)$.} \label{fig9}
\end{figure}

The reconstruction is theoretically ill posed.  Worse, the kernel
functions, shown in Figure~\ref{fig8}, have something in common; they
all have the same value, and zero slope, at the point $\omega=0$,
exactly where we want to know our function.  Therefore, if there is
nontrivial structure in $\sigma(\omega)$ close to $\omega=0$, we will
miss it.  Perturbatively, there is such structure.  Nonperturbatively,
we just don't know.  In an analog theory, $\mathcal{N}=4$ SYM theory,
we know that there is \textsl{no} such structure \cite{Teaney}; yet the
reconstructed $\sigma(\omega)$ is almost the same as in weak-coupling
QCD, as illustrated in Figure~\ref{fig9}.

\section{Conclusions}
\label{sec:conclusion}

Shear viscosity is an important property of QCD, which plays a central
role in its dynamics.  Computing shear viscosity from first principles
is generally hard, and it seems to be especially hard in the theory of
QCD.  We showed how the perturbative approach encounters challenges
associated with diagrammatic resummation, and results in a poorly
convergent perturbation theory.  The prospects of determining QCD on
the lattice are also challenged by the need to reconstruct a Minkowski
function from its Euclidean continuation with error bars.  In known
examples, very distinct Minkowski functions arise from nearly
identical Euclidean continuations.  Nevertheless, we see this as the
second-best approach to determine shear viscosity for QCD at this
time.  The best approach is to try to extract it from data.  This is
also challenging but a different speaker should address this issue.

\bibliographystyle{JHEP}
\bibliography{refs.bib}

\end{document}